\documentclass{PoS}

\usepackage{amsmath}
\usepackage{slashed}

\usepackage{calrsfs}
\DeclareMathAlphabet{\pazocal}{OMS}{zplm}{m}{n}
\usepackage{newtxtext,newtxmath}

\newcommand{\tr}{{\textrm {tr}}}

\newcommand{\U}{{\textrm {U}}}

\newcommand{\cB}{{\pazocal B}}
\newcommand{\cV}{{\pazocal V}}
\newcommand{\DC}{{\textrm {DC}}}
\newcommand{\tot}{{\textrm {tot}}}
\newcommand{\CS}{{\textrm {CS}}}
\newcommand{\CSK}{{\textrm {CSK}}}
\newcommand{\matter}{{\textrm {matter}}}
\newcommand{\GH}{{\textrm {GH}}}

\newcommand{\J}{{\pazocal{J}}}

\title{Anomalous transport, massive gravity theories and holographic momentum relaxation}

\ShortTitle{Anomalous transport, massive gravity theories and holographic momentum relaxation}

\author{\speaker{Eugenio Meg\'{\i}as}$\;^{a,b}\,$   \\
        \llap{$^a$} Departamento de F\'{\i}sica At\'omica, Molecular y Nuclear and \\ Instituto Carlos I de F\'{\i}sica Te\'orica y Computacional, Universidad de Granada, \\ Avenida de Fuente Nueva s/n,  18071 Granada, Spain \\
        \llap{$^b$} Departamento de F\'{\i}sica Te\'orica, Universidad del Pa\'{\i}s Vasco UPV/EHU, \\ Apartado 644, 48080 Bilbao, Spain \\
      E-mail: \email{emegias@ugr.es}}

\abstract{Quantum anomalies give rise to new non-dissipative transport
  phenomena in relativistic fluids induced by external electromagnetic
  fields and vortices. These phenomena can be studied in holographic
  models with Chern-Simons couplings dual to anomalies in field
  theory. We perform a computation in AdS/CFT of the anomalous
  transport coefficients in a holographic massive gravity model, and
  find that the anomalous conductivities turn out to be independent of
  the holographic disorder couplings of the model. To arrive at this
  result we suggest a new definition of the energy-momentum tensor in
  presence of the gauge-gravitational Chern-Simons coupling. We also
  compute the electric DC conductivity and find that it can vanish for
  certain values of the disorder couplings.}

\FullConference{XIII Quark Confinement and the Hadron Spectrum - Confinement2018\\
		31 July - 6 August 2018\\
		Maynooth University, Ireland}

\begin{document}

\section{Introduction}
\label{sec:introduction}

The modern understanding of hydrodynamics is as an effective field theory~\cite{Kovtun:2012rj}. The equations of motion of the hydrodynamical systems correspond to the conservation laws of the energy-momentum tensor and charged currents. In presence of quantum anomalies, however, the currents are no longer conserved, and this has important consequences for the transport properties of the systems. This is the case of gauge anomalies, which are responsible for new  dissipationless transport phenomena, such as the {\it chiral magnetic} and {\it chiral vortical} effects (see e.g.~\cite{Landsteiner:2012kd,Kharzeev:2013ffa} for some reviews). In addition to anomalies, there are other sources of non-conservation of the currents, such as {\it disorder} effects in condensed matter systems, see e.g.~\cite{Lee:1985zzc}. From a field theory perspective, disorder is related to explicit breaking of translational invariance, and this leads to momentum dissipation and a modification of the conservation law of the energy-momentum tensor. A possible way of studying this in holography is by giving a mass to the graviton. A particular realization of this idea is based on the introduction of massless scalar fields in AdS space with spatially linear profiles~\cite{Andrade:2013gsa}. Some dissipative transport coefficients, like the DC electric conductivity, have been studied with these holographic massive gravity models leading to a decreasing value when disorder increases~\cite{Gouteraux:2016wxj,Baggioli:2016oqk}. In this work we will address the generalization of these models to study non-dissipative transport properties. A reanalysis of the form of the holographic energy-momentum tensor will be needed to get a consistent result.

\section{Hydrodynamics of relativistic fluids}
\label{sec:Hydro}

The basic ingredients to study hydrodynamics are the (anomalous) conservations laws of the energy-momentum tensor and charged currents. These are supplemented by the constitutive relations, i.e. expressions of the energy-momentum tensor and the currents in terms of fluid quantities, which are organized in a derivative expansion, also called hydrodynamic expansion~\cite{Kovtun:2012rj},
\begin{eqnarray}
\langle T^{\mu\nu} \rangle &=& (\varepsilon + P) u^\mu u^\nu  + P g^{\mu\nu} + (\sigma_{\varepsilon}^{\cB})_a (B_a^\mu u^\nu + B_a^\nu u^\mu)  + \sigma_{\varepsilon}^{\cV} (\Omega^\mu u^\nu +  \Omega^\nu u^\mu) + \cdots \,,  \label{eq:T_hydro}\\
\langle J_a^\mu \rangle &=& n_a u^\mu + \sigma_{ab} \left( E_b^\mu - T P^{\mu\nu} D_\nu \left( \frac{\mu_b}{T} \right) \right)  +  \sigma_{ab}^\cB B_b^\mu + \sigma_a^\cV\Omega^\mu + \cdots \,.  \label{eq:J_hydro}
\end{eqnarray}
Here $\varepsilon$ is the energy density, $P$ is the pressure, $n_a$ are the charge densities, $u^\mu$ is the local fluid velocity and $P^{\mu\nu} = g^{\mu\nu} + u^\mu u^\nu$ is the transverse projector to the fluid velocity. External electric and magnetic fields are covariantized as $E_a^\mu = F_a^{\mu\nu} u_\nu$ and $B_a^\mu =  \frac{1}{2} \epsilon^{\mu\nu\rho\sigma} u_\nu F_{a,\rho\sigma}$, where the field strengths of the gauge fields in an Abelian theory are defined as $F_{a,\mu\nu} = \partial_\mu A_{a,\nu} - \partial_\nu A_{a,\mu}$. In addition to the equilibrium contributions, there are extra terms in the constitutive relations which lead to dissipative and anomalous transport effects. While the electric conductivities~$\sigma_{ab}$ in Eq.~(\ref{eq:J_hydro}) are responsible for dissipative transport, we find in this equation two examples of anomalous transport: i) the {\it chiral magnetic effect}, which is responsible for the generation of an electric current parallel to a magnetic field~\cite{Kharzeev:2009pj}, and ii) the {\it chiral vortical effect} in which the electric current is induced by a vortex in the fluid~$\Omega^\mu = \epsilon^{\mu\nu\rho\sigma} u_\nu \partial_\rho u_\sigma$~\cite{Son:2009tf,Landsteiner:2012kd}. Apart from the charge flow in a relativistic fluid, there exists also energy flow and consequently analogous anomaly related transport effects in the energy current $J_\varepsilon^i \equiv \langle T^{0i} \rangle$, cf. Eq.~(\ref{eq:T_hydro}).~\footnote{At first order in derivatives the notion of fluid velocity is ambiguous, and needs to be fixed by prescribing a choice of frame. Here we choose a frame in which we demand that the definition of the fluid velocity is not influenced when switching on an external magnetic field or having a vortex.}

A convenient way to compute the anomalous conductivities are the Kubo formulae, which are based on retarded correlators of the charged currents and the energy-momentum tensor, and they are obtained within linear response theory~\cite{Landsteiner:2012kd}. Using this formalism for a theory of free chiral fermions, it has been found that the 1-loop calculation of the chiral magnetic and vortical conductivities receive contributions of the axial anomaly~\cite{Kharzeev:2009pj}, and also of the mixed gauge-gravitational anomaly~\cite{Landsteiner:2011cp}. More explicitly, the conductivities read
\begin{equation}
\sigma_{ab}^\cB =  \frac{1}{4\pi^2} d_{abc} \mu^c  \,, \qquad \sigma_a^\cV =  \frac{1}{8\pi^2}  d_{abc} \, \mu^b \, \mu^c + \frac{T^2}{24} b_a \,,
\end{equation}
where $d_{abc} = \frac{1}{2} \tr(T_a \{T_b, T_c\})_L - \frac{1}{2}
\tr(T_a \{T_b, T_c\})_R$ and $b_a = \tr(T_a)_L - \tr(T_a)_R$ are the
group theoretic factors related to the axial and gauge-gravitational
anomalies, respectively. The Kubo formulae also predict that the
chiral vortical conductivity coincides with the chiral magnetic
conductivity for the energy current~$\sigma_a^\cV =
(\sigma^\cB_\varepsilon)_a$. Apart from the Kubo formalism, the
anomalous transport coefficients have been studied in a wide variety
of methods, either in field theory or in holography, leading to
similar results: these include diagrammatic
methods~\cite{Manes:2012hf}, fluid/gravity
correspondence~\cite{Erdmenger:2008rm,Landsteiner:2011iq}, and the
partition function
formalism~\cite{Banerjee:2012iz,Jensen:2012jy,Megias:2014mba}. In this
work we will study anomalous transport phenomena in a holographic
massive gravity model in the context of linear response theory. The
electric DC conductivity will be computed as well in this model, and
compared with previous studies. We will deal with a single $\U(1)$
symmetry.

\section{Massive gravity and holographic momentum relaxation}
\label{sec:massive_gravity}

In massive gravity theories, the momentum relaxation is described through the St\"uckelberg mechanism with Goldstone modes corresponding to scalar fields, $X^I$, which are related to spatial translations. A recent holographic implementation of this idea in 4-dim has been presented in Refs.~\cite{Gouteraux:2016wxj,Baggioli:2016oqk}. In this work we will consider explicitly the model of~\cite{Gouteraux:2016wxj}, but the same conclusions are obtained when considering the model of~\cite{Baggioli:2016oqk}.

\subsection{The model}
\label{subsec:model}

In order to study anomalous transport in holographic massive gravity theories, one should consider the theory in odd dimensions, as only in this case one can introduce in the action the corresponding Chern-Simons (CS) terms that account for the effects of quantum anomalies. For the moment we will focus on non-anomalous properties, and leave the study of the CS terms for Sec.~\ref{sec:anomalous_transport}. The action of the model in 5-dim is
\begin{equation}
 S = \int d^5x \sqrt{-g} \left[ R + 12 - {\frac 1 2 \partial^M X^I \partial_M X^I} - \frac 1 4 F^2 - {\frac{\J}{4} \partial_M X^I \partial_N X^I F^N\,_L F^{L M} } \right] + S_{\GH}  \,, \label{eq:S}
\end{equation}
with scalars $X^I = k \, \delta^I_i \, x^i$ that break translational invariance.~\footnote{In the following we will adopt the following notation: capital letters '$M$' denote 5-dim indexes, greek letters '$\mu$' denote 4-dim indexes in the holographic boundary of AdS, and lower-case latin letters '$i$' denote spatial directions in that boundary.} The parameter $k$ controls the degree of breaking of translational invariance, and in particular when $k=0$ one recovers the massless gravity theory. The coupling~$\J$ represents the effects of disorder on the charged sector of the theory. $S_{\GH}$ is the usual Gibbons-Hawking boundary term. In the following we will consider as background a charged black hole solution with AdS asymptotics of the form
\begin{equation}
ds^2 = \frac{1}{u}\left(-f(u) dt^2 + dx^2+dy^2+dz^2\right) + \frac{du^2}{4u^2f(u)}\,, \qquad A_t = \phi(u)\,. 
\end{equation}
The solutions of the equations of motion are then
\begin{equation}
f(u) = \left(1 - \frac{u}{u_h}\right) \left( 1 + \frac{u}{u_h} - \frac{k^2}{4}u - \frac{\mu^2}{3} \frac{u^2}{u_h} \right) \,, \qquad \phi(u) = \mu  \left( 1 - \frac{u}{u_h} \right) \,,
\end{equation}
where we identify $\mu$ with the chemical potential. The temperature~$T = \frac{1}{\pi\sqrt{u_h}} \left( 1 - \frac{k^2}{8} u_h - \frac{\mu^2}{6} u_h \right)$ does not depend on the charge disorder coupling~$\J$.

\subsection{Electric DC conductivity}
\label{sec:DC_conductivity}

\begin{figure*}[htb]
\begin{tabular}{cc}
\includegraphics[width=70mm]{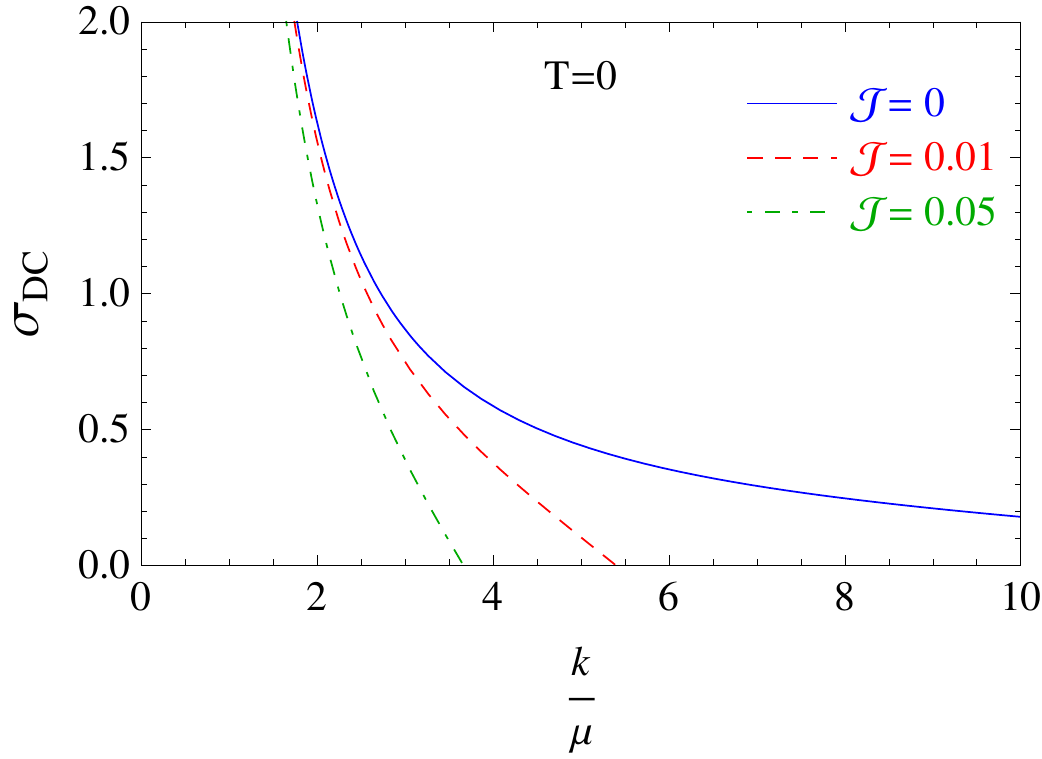} & \hspace{1cm}
\includegraphics[width=60mm]{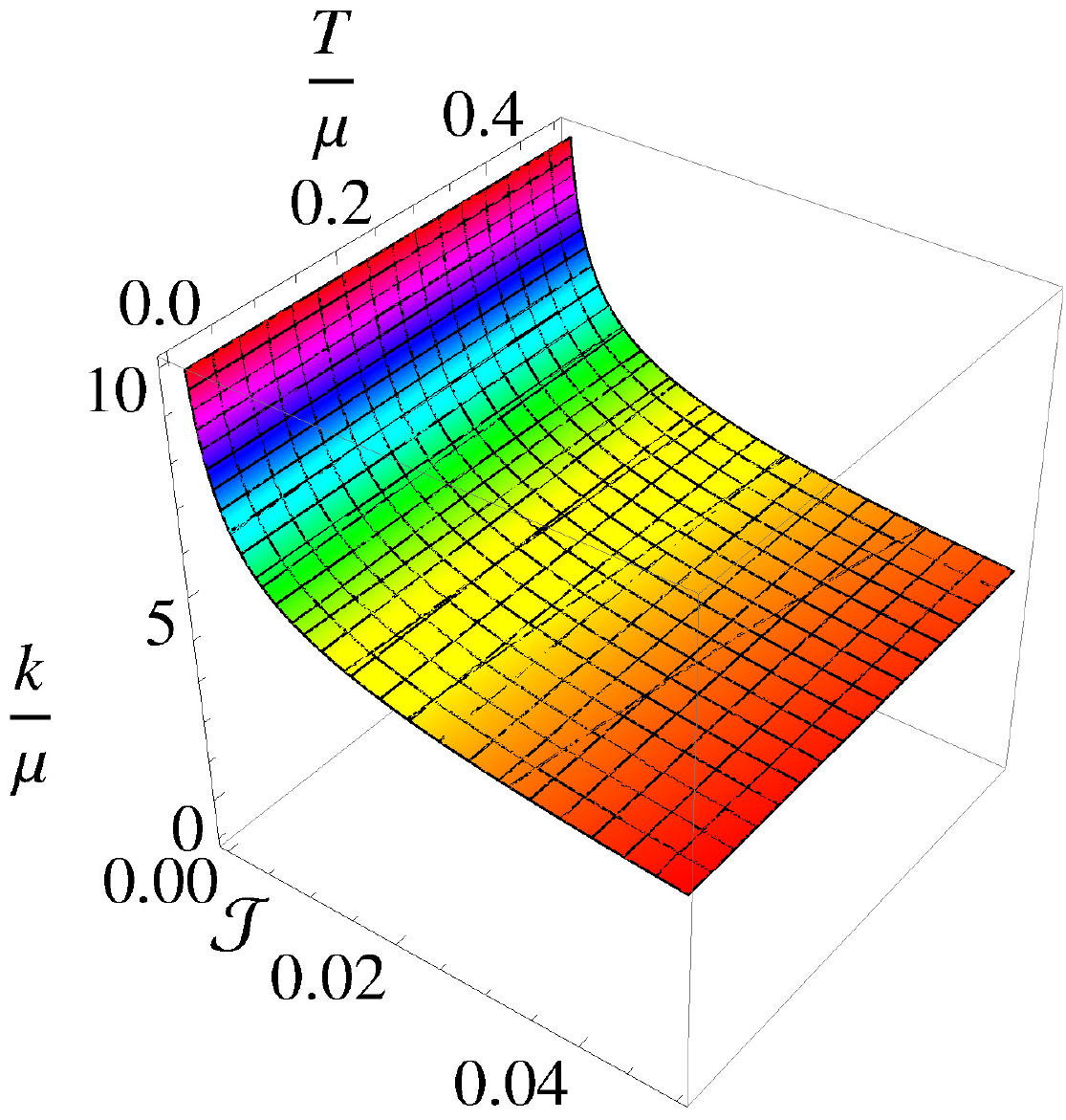} \\
\end{tabular}
\caption{Left panel: DC conductivity at zero temperature as a function of $k$ (normalized to the chemical potential $\mu$). Right panel: Region in the plane $(\J,T/\mu,k/\mu)$ where the DC conductivity of Eq.~(\ref{eq:sigmaDC}) 
vanishes.}
\label{fig:DC_conductivity}
\end{figure*}

The electric conductivity measures the electric current $J^\mu$ induced by an electric field $E^\mu$, cf. Eq.~(\ref{eq:J_hydro}). There are several methods to compute the DC conductivity in holography, but one of the most straightforward is the one proposed in Ref.~\cite{Donos:2014cya} within linear response theory. Let us consider the small perturbations in the metric and gauge field
\begin{equation}
A_z = \epsilon (-E t + a_z(u)) \,, \qquad g_{tz} = \frac{\epsilon}{u} h^z_t(u) \,, \qquad g_{uz} = \frac{\epsilon}{u} h^z_u(u) \,,
\end{equation}
where $E$ is the external electric field in the $z$-direction, and assume that the perturbations do not introduce additional sources, i.e. $a_z(0) = h^z_t(0) = h^z_u(0) = 0$. The equations of motion for the perturbations can be solved by demanding regularity of the metric. Noting that the electric current in the $z$-direction is $J_z = 2 \lim_{u \to 0}(f(u) a_z^\prime(u))$, and without going into the details of the solution, we obtain the DC conductivity~\cite{Megias:2016aje,Copetti:2017ywz} 
\begin{equation}
  \sigma_{\DC} = \frac{J_z}{E}= \frac{1}{\sqrt{u_h}} \left(1 - \frac{k^2}{2} \J u_h \right) \left[1 + \left(1 - \frac{k^2}{2} \J  u_h \right) \frac{ 4 \mu^2}{k^2 (1 + 2 \J \mu^2 u_h)} \right] \,.  \label{eq:sigmaDC}
\end{equation}
Note that in the case $\J = 0$, the DC conductivity is $\sigma_{\DC} =  (1 + 4\mu^2/k^2) / \sqrt{u_h} > 1$, so that it is bounded from below.~\footnote{This bound is similar to the one proven for holographic matter in $(2+1)$-dim in Ref.~\cite{Grozdanov:2015qia}.} On the other hand, the conductivity vanishes for $k^2 \J u_h = 2$, so that in this regime the system behaves as an insulator. Moreover, $\sigma_{\DC}$~can become even negative in some range of the parameters indicating an instability. We show in Fig.~\ref{fig:DC_conductivity}~(left) the value of the DC conductivity as a function of the parameter~$k$ for different values of the parameter~$\J$. We also display in Fig.~\ref{fig:DC_conductivity}~(right) the regime of parameters where the DC conductivity vanishes.

\section{Anomalous transport in massive gravity theories}
\label{sec:anomalous_transport}

To study the non-dissipative transport properties of the 5-dim massive gravity model, we should introduce anomalous effects in the theory. As mentioned above, anomalies are mimicked in the gravity side through CS terms in the action. Then, we can extend the model of Eq.~(\ref{eq:S}) by adding the following gauge and mixed gauge-gravitational CS terms~\cite{Landsteiner:2011iq,Megias:2013joa}
\begin{equation}
S_\mathrm{CS} = \int d^5x \sqrt{-g} \epsilon^{\mu\nu\rho\sigma\tau} A_\mu \left( \frac{\kappa}{3} F_{\nu\rho}F_{\sigma\tau} + \lambda R^\alpha\,_{\beta\nu\rho}R^\beta\,_{\alpha\sigma\tau}\right) \,. \label{eq:S_CS}
\end{equation}
and 
\begin{equation}
S_{\CSK} = -8 \lambda \int_\partial d^4x \sqrt{\gamma} n_\mu \epsilon^{\mu\nu\rho\sigma\tau} A_\nu K_{\rho\beta} D_\sigma K^\beta_{\tau} \,, \label{eq:S_CSK}
\end{equation}
so that the total action is $S_{\tot}  = S + S_{\CS} + S_{\CSK}$. Eq.~(\ref{eq:S_CSK}) is a convenient counterterm that allows to reproduce the gravitational anomaly at general hypersurface, $\gamma_{\mu\nu}$ is the induced metric and $K_{\mu\nu}$ is the extrinsic curvature on the holographic boundary of an asymptotically AdS space defined by an outward pointing unit normal vector~$n^\mu$.

\subsection{Holographic energy-momentum tensor}
\label{subsec:holographic_EMT}

Let us consider for the moment the standard Fefferman-Graham coordinates,~$ds^2 = dr^2 + \gamma_{\mu\nu} dx^\mu dx^\nu$. The variation of the on-shell action with a timelike hypersurface at a fixed $r$ is 
\begin{equation}
\delta S_{\tot} = \frac{1}{2}\int_\partial \sqrt{-\gamma} \left( t^{\mu\nu} \delta \gamma_{\mu\nu} + u^{\mu\nu}\delta K_{\mu\nu} \right) + \delta S_{\matter} \,,
\end{equation}
where we keep $\gamma_{\mu\nu}$ and $K_{\mu\nu}$ as independent variables, i.e. the extrinsic curvature acts like an external source conjugate to the operator~$u^{\mu\nu}$. We will define the holographic energy-momentum tensor as~\cite{Copetti:2017ywz,Copetti:2017cin}
\begin{equation}
{T^\mu}_\nu =  t^\mu\,_\nu + u^{\mu\rho} K_{\rho\nu} \,, \qquad \textrm{with} \qquad t^{\mu\nu} = t_0^{\mu\nu} + t_\lambda^{\mu\nu} \,,  \label{eq:emtensornew}
\end{equation}
where $t_0^{\mu\nu} =  - 2 \sqrt{-\gamma} ( K^{\mu\nu} - K \gamma^{\mu\nu} )$ is the standard Brown-York contribution, and
\begin{align}
 t_\lambda^{\mu\nu} &=- 8\lambda \sqrt{-\gamma} \epsilon^{\rho\sigma\tau(\mu} \left( 2 D_\sigma K_\tau^{\nu)} F_{r\rho} + \gamma^{\nu)\beta} \dot{K}_{\beta\sigma} F_{\tau\rho} - F_{\tau\rho} K_\beta^{\nu)} K_\sigma^\beta \right) \, , \\
 u^{\mu\nu} &= 8 \lambda \sqrt{-\gamma} \epsilon^{\rho\sigma\tau(\mu} F_{\rho\sigma} K_\tau^{\nu)} \,,
\end{align}
where $A_{(\mu\nu)} := \frac{1}{2}(A_{\mu\nu} + A_{\nu\mu})$, and dot denotes differentiation with respect to~$r$. This result naturally follows from the Ward identity of the energy-momentum tensor in presence of the gauge-gravitational CS term~\cite{Copetti:2017ywz}.~\footnote{In the case of holographic pure gravitational anomalies dual to 2-dim field theories a similar correction has been found in Ref.~\cite{Kraus:2005zm}.} While ${(t_0)^\mu}_\nu$ is divergent and needs to be regularized by the standard counterterms~\cite{deHaro:2000vlm}, the contributions ${(t_\lambda)^\mu}_\nu$ and $u^{\mu\rho}K_{\rho\nu}$ are already finite before the holographic renormalization is performed. As we will see in the following, the extra contributions ${(t_\lambda)^\mu}_\nu$ and $u^{\mu\rho}K_{\rho\nu}$ are essential to get the physically correct results for the anomalous transport coefficients.

\subsection{Chiral magnetic conductivity}
\label{subsec:CMC}

To compute this conductivity in linear response theory, we consider the perturbation
\begin{equation}
A_y = \epsilon B x \,, \qquad A_z = \epsilon a_z(u) \,, \qquad g_{tz} = \frac{\epsilon}{u} h_t^z(u) \,.
\end{equation}
If we apply the usual holographic dictionary to the solution of the equations of motion of the perturbations, and compute the energy-momentum tensor from these solutions, one finds~$T_{0z} = 4 g^\prime_{tz}(u=0) = \left(\kappa 4 \mu^2 + \lambda 32 \pi^2 T^2  - \lambda 2 k^2 \right) B$~\cite{deHaro:2000vlm}, which corresponds only to the contribution $(t_0)^{\mu\nu}$. Using the background, one can easily calculate the corrections to the energy-momentum tensor due to $(t_\lambda)^{\mu\nu}$ and $u^{\mu\rho} K_{\rho\nu}$, leading to~$(t_\lambda)^{\mu\nu} = 0$ and $u^{\mu\rho} K_{\rho}^{\nu} = \lambda 4 k^2 B \delta^{\mu(0} \delta^{z)\nu}$. Finally, from the new definition of the energy-momentum tensor given by Eq.~(\ref{eq:emtensornew}), we find
\begin{equation}
\vec{J} = \kappa 8 \mu \vec{B} \,, \qquad \vec{J}_\varepsilon = \left(\kappa 4 \mu^2 + \lambda 32 \pi^2 T^2 \right) \vec{B} \,,
\end{equation}
which are the usual expressions for the chiral magnetic effect in the charge and energy currents~\cite{Landsteiner:2012kd}.

\subsection{Chiral vortical conductivity}
\label{subsec:CVC}
 
Vorticity $\Omega^i = \epsilon^{ijk} \partial_j u_k$ can be introduced through a gravitomagnetic field $B_g^i = \epsilon^{ijk} \partial_j (A_g)_k$. In the rest frame $u^\mu = (1,0,0,0)$, the gravitomagnetic field is in the mixed component of the metric, i.e. $ds^2 = -dt^2 + 2 (A_g)_i dt dx^i + d\vec{x}^2$, and it follows~$\vec{B}_g = \vec{\Omega}$~\cite{Landsteiner:2011tg}. Let us consider the ansatz
\begin{equation}
A_y = \epsilon B_g u \mu x \,, \qquad A_z = \epsilon a_z(u) \,, \qquad g_{ty} = \epsilon \frac{f(u)}{u} B_g x \,, \qquad g_{tz} = \frac{\epsilon}{u} h_t^z(u) \,.
\end{equation}
After solving the equations of motion for the perturbations, one finds from the $u \to 0$ asymptotics of $a_z(u)$ and $h_t^z(u)$ the following response due to a gravitomagnetic field
\begin{equation}
\vec{J} = \left(\kappa 4 \mu^2 + \lambda 32\pi^2 T^2 \right) \vec{B}_g \,, \qquad \vec{J}_\varepsilon = \left( \kappa \frac{8}{3} \mu^3 + \lambda 64 \pi^2 T^2\mu \right) \vec{B}_g \,.
\end{equation}
Remembering that $\vec{B}_g = \vec{\Omega}$, these are the usual responses of a chiral fluid due to vorticity.

\subsection{Discussion on anomalous transport}
\label{subsec:discussion_anomalous_transport}

The results of this section can be summarized in the following values for the conductivities~\cite{Copetti:2017ywz}~\footnote{To compare to a free theory of $N_f$ chiral fermions we can identify $\kappa = N_f/(32\pi^2)$ and $\lambda = N_f/(768\pi^2)$.}
\begin{equation}
  \sigma^\cB          =  \kappa 8\mu \,, \hspace{2.88cm} \sigma^\cV          =  \kappa 4\mu^2 + \lambda 32\pi^2 T^2  \,, 
\end{equation}

\begin{equation}
  \sigma_\varepsilon^\cB   =  \kappa 4\mu^2 + \lambda 32 \pi^2 T^2 \,, \hspace{1cm} \sigma_\varepsilon^\cV   =  \kappa \frac{8}{3}\mu^3 + \lambda 64\pi^2 T^2 \mu \,.
\end{equation}
Note that these values are the same as in massless gravity, i.e. they don't have any dependence on the holographic disorder couplings $(k,\J)$. This means that the anomalous conductivities are not affected by translational breaking effects, and this constitutes one of the most important results of this work. From this property, together with the result of Sec.~\ref{sec:DC_conductivity}, we conclude that there is a regime in the theory in which the DC conductivity vanishes, but the anomalous conductivities do not vanish. Finally, one important aspect to remark is that the equality between $\sigma_\varepsilon^\cB$ and $\sigma^\cV$ follows non-trivially from the definition of the energy-momentum tensor in Eq.~(\ref{eq:emtensornew}), as the term $u^{\mu\rho} K_{\rho\nu}$ induces a contribution which exactly cancels a dependence $ \sigma_\varepsilon^\cB \propto \lambda k^2$~\cite{Copetti:2017ywz}.

\section{Conclusions}
\label{sec:conclusions}

Massive gravity theories have been introduced in the literature as holographic duals of disorder in condensed matter systems. In this work we have studied non-dissipative transport properties  induced by external electromagnetic fields and vortices in these theories, in particular the chiral magnetic and chiral vortical effects. We found, as expected, that the corresponding conductivities are unchanged by the holographic disorder effects. This property, for the case of the response in the energy current, follows non-trivially from a careful study of the energy-momentum tensor, which turns out to be modified by the presence of the mixed gauge-gravitational Chern-Simons term in the action~\cite{Copetti:2017ywz}. This solves the puzzle found in Ref.~\cite{Megias:2016aje}. In addition, we have studied the electric DC conductivity, and found an interesting regime in which it vanishes, but the anomalous conductivities do not vanish. This leads to the possibility of studying the anomalous transport effects of these systems in this regime in a clean way.

\acknowledgments

This work is based on Refs.~\cite{Megias:2016aje,Copetti:2017ywz}. I
would like to thank J.~Fern\'andez-Pend\'as and especially
K.~Landsteiner for enlightening discussions.  Research supported by
the Spanish MINEICO and European FEDER funds grants
FPA2015-64041-C2-1-P and FIS2017-85053-C2-1-P, by the Junta de
Andaluc\'{\i}a grant FQM-225, and by the Basque Government grant
IT979-16. The research of E.M. is also supported by the Ram\'on y
Cajal Program of the Spanish MINEICO, and by the Universidad del
Pa\'{\i}s Vasco UPV/EHU, Bilbao, Spain, as a Visiting Professor.


\end{document}